\documentclass{Interspeech}

\interspeechcameraready

\title{Leveraging Cascaded Binary Classification and Multimodal Fusion for Dementia Detection through Spontaneous Speech\thanks{* Corresponding author.}}
\author[affiliation={1,4}]{Yin-Long}{Liu}
\author[affiliation={3}]{Yuanchao}{Li}
\author[affiliation={1}]{Rui}{Feng}
\author[affiliation={2}]{Liu}{He}
\author[affiliation={1,4}]{Jia-Xin}{Chen}
\author[affiliation={1}]{Yi-Ming}{Wang}
\author[affiliation={4}]{Yu-Ang}{Chen}
\author[affiliation={4}]{Yan-Han}{Peng}
\author[affiliation={1,2*}]{Jia-Hong}{Yuan}
\author[affiliation={1,2,4}]{Zhen-Hua}{Ling}

\affiliation{National Engineering Research Center of Speech and Language Information Processing}{University of Science and Technology of China}{Hefei, P. R. China}
\affiliation{Interdisciplinary Research Center for Linguistic Sciences}{University of Science and Technology of China}{Hefei, P. R. China}
\affiliation{Centre for Speech Technology Research}{University of Edinburgh}{UK}
\affiliation{Department of Electronic Engineering and Information Science}{University of Science and Technology of China}{Hefei, P. R. China}
\email{lyl2001@mail.ustc.edu.cn, \{jiahongyuan, zhling\}@ustc.edu.cn}


\usepackage{comment}

\usepackage{cite}
\usepackage{hyperref}
\usepackage{algorithmic}
\usepackage{textcomp}
\usepackage{xcolor}
\usepackage{enumerate}
\usepackage{cleveref}
\usepackage{graphicx}    
\usepackage{subcaption} 
\usepackage{makecell}
\usepackage{enumitem}
\usepackage[bottom]{footmisc} 
\usepackage{bm}
\usepackage{adjustbox}
\usepackage{amsmath,amsthm,amssymb,amsfonts}
\usepackage{bigstrut,multirow,rotating,booktabs} 

\begin{document}

\maketitle
% \vspace*{1.2cm} % 尝试使用 \vspace*
\begin{abstract}
% Early detection of dementia, such as mild cognitive impairment, remains a critical challenge in dementia research. 
This paper presents our submission to the PROCESS Challenge 2025, focusing on spontaneous speech analysis for early dementia detection. For the three-class classification task (Healthy Control, Mild Cognitive Impairment, and Dementia), we propose a cascaded binary classification framework that fine-tunes pre-trained language models and incorporates pause encoding to better capture disfluencies. This design streamlines multi-class classification and addresses class imbalance by restructuring the decision process. For the Mini-Mental State Examination score regression task, we develop an enhanced multimodal fusion system that combines diverse acoustic and linguistic features. Separate regression models are trained on individual feature sets, with ensemble learning applied through score averaging. Experimental results on the test set outperform the baselines provided by the organizers in both tasks, demonstrating the robustness and effectiveness of our approach.

\end{abstract}
\begin{figure*}[t!]
    \centering
\includegraphics[width=1\linewidth]{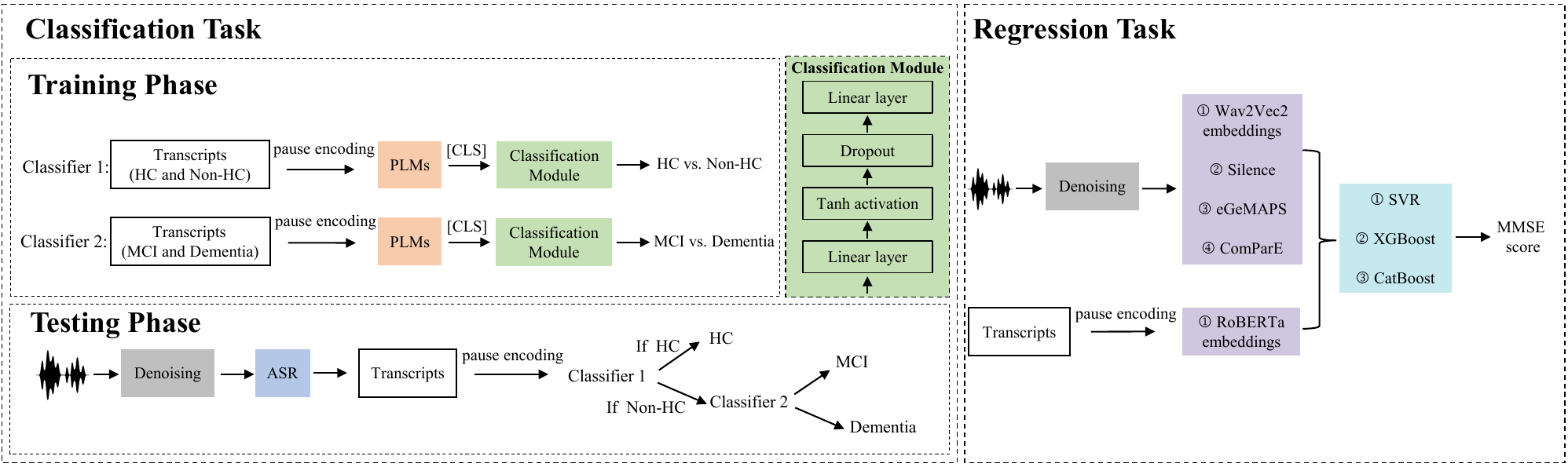}
% \vspace{-0.6cm}
    \caption{Architecture of methods for classification and regression tasks. PLMs refer to the pre-trained language models BERT, RoBERTa, and ERNIE. Non-HC refers to a label that is either MCI or Dementia. For the training phase, we use manual transcripts, while during the testing phase, ASR transcripts are used. The regression task shown in the figure represents the training phase.}
    \label{fig:Over workflow.}
\vspace{-0.6cm}
\end{figure*}

\section{Introduction}
Dementia is a neurodegenerative disease that worsens over time and causes irreversible damage to the brain, manifested by a persistent deterioration of an individual's cognitive and functional abilities, including memory, attention, and executive function \cite{beck1982dementia}.
Mild Cognitive Impairment (MCI), an early dementia symptom, is typically regarded as a transitional stage between normal aging and dementia, leading to a mild decline in cognitive functions without significantly impairing daily activities.
% According to the World Health Organization, more than 50 million people worldwide are currently living with dementia, with millions of newly diagnosed cases each year, which could increase to more than 152 million by 2050 \cite{nichols2022estimation}. Studies have shown that mid-course intervention before neuronal degeneration in the brain can effectively alleviate the problem, highlighting the critical importance of early dementia detection \cite{nestor2004advances}.
Today, several biochemical methods, such as magnetic resonance imaging and positron emission tomography, have been developed for early dementia evaluation \cite{jack2008alzheimer,samper2018reproducible} . However, these methods are often expensive, demanding, time-consuming, and require validation by neurologists in manual clinical settings. Therefore, more efficient and reliable automated methods are needed to achieve early dementia detection \cite{li2024semi}. While memory impairment is widely recognized as the hallmark symptom of dementia, speech and language impairments are equally prevalent and serve as critical indicators for the early detection of the disease \cite{cummings1989speech,yeung2021correlating}. Compared to traditional clinical methods, speech- and language-based automatic dementia diagnosis 
has garnered increasing attention due to its non-invasive, cost-effective, and convenient nature \cite{talkar24_interspeech,liu24f_interspeech,li2022alzheimer,pan2021using,liu2025can}. PROCESS (Prediction and Recognition Of Cognitive declinE through Spontaneous Speech) Signal Processing Grand Challenge \cite{tao2024early} focuses on early dementia detection through the analysis of spontaneous speech. This challenge consists of two tasks: a classification task and a regression task. Unlike the binary classification tasks in previous challenges \cite{luz2020fuente,luz2021detecting,luz2023multilingual}, PROCESS involves a three-class classification task, aiming to predict the diagnostic label of the subjects as Healthy Control (HC), MCI, or Dementia. The inclusion of the MCI category increases the difficulty of the classification task. The regression task is similar to those in previous challenges, aiming to predict the Mini-Mental State Examination (MMSE) score reflecting cognitive decline. In this work, we propose:
\begin{itemize}
    \item A cascaded binary classification framework that incorporates pause encoding to effectively address multi-class challenges.
    \item An enhanced multimodal fusion system that seamlessly integrates diverse acoustic and linguistic features for improved performance.
\end{itemize}

Experimental results, in contrast to the baseline, validate the efficacy of our approach and highlight its potential in multi-class classification and early detection of dementia.

\vspace{-0.2cm}
\section{Related Work}
\vspace{-0.2cm}
Current research primarily utilizes two types of features extracted from spontaneous speech for dementia detection: acoustic features from raw audio signals and linguistic features from transcripts \cite{pan2021using, luz2021detecting,braun24_interspeech}. Both features offer unique insights into cognitive decline associated with Dementia. Based on the recent studies \cite{mei2023ustc,syed2020automated,syed2021tackling,rohanian2021alzheimer}, acoustic features used in dementia detection can be divided into prosody, duration of pauses, emotional embeddings, pre-trained models embeddings features, and more. Key linguistic features include, but are not limited to, lexical richness, syntactic complexity and pre-trained textual embedding features \cite{martinc2020tackling,qiao2021alzheimer,santos2017enriching}. Previous research has shown that, compared to using acoustic features alone, employing linguistic features
extracted from transcripts is more effective in distinguishing between Dementia and HC \cite{syed2020automated,li2021comparative}. Advanced Natural Language Processing (NLP) techniques, particularly those involving Pre-trained Language Models (PLMs) such as BERT \cite{devlin2018bert}, RoBERTa \cite{liu2019roberta}, and ERNIE \cite{sun2020ernie}, are capable of capturing complex patterns and contextual information that are crucial for detecting cognitive decline, and have proven effective in dementia detection \cite{wang22l_interspeech, yuan2020disfluencies}. 

Nevertheless, very limited research has explored the three-class classification problem in dementia detection, posing challenges in the early diagnosis of dementia, particularly MCI. To address this long-standing issue, we propose a cascaded binary classification approach. Furthermore, we investigate the integration of multimodal features to enhance performance in the regression task. 

\vspace{-0.3cm}
\section{Dataset}
\vspace{-0.2cm}
\label{sec:Dataset}
The provided dataset contains a total of 157 subjects, later divided into training and validation sets. Each subject has a diagnostic label (82 HC, 59 MCI, 16 Dementia) for the classification task. For the regression task, MMSE scores are available for a subset of 69 subjects. To evaluate language communication and cognitive function, subjects completed three standardized English elicitation tasks commonly used in dementia research: the Cookie Theft Description (CTD), Phonemic Fluency Test (PFT), and Semantic Fluency Test (SFT).

For the CTD task, subjects describe the ``Cookie Theft'' picture, originally part of the Boston Diagnostic Aphasia Examination and later widely used for dementia detection. This task evaluates cognitive functions, including language comprehension and memory. In the PFT task, subjects list as many words as possible starting with the letter ``P'' within one minute, excluding proper nouns and country names, assessing verbal fluency and executive language functions. The SFT task requires subjects to name as many animals as possible in one minute, primarily evaluating naming skills and language comprehension to identify potential language impairments.

Each of the 157 subjects has an audio recording and a corresponding manual transcript for each of the elicitation tasks described above. The manual transcripts include annotations for speaker identities, external noises, and durations between utterances. To align the transcripts with the actual spoken content, we remove these annotations. For the test set, only the audio recordings are available, while manual transcripts and diagnostic labels are withheld.

The dataset presents a significant class imbalance, with a predominance of HC subjects and relatively few Dementia cases. While this mirrors real-world clinical distributions, it complicates the training of classification models and motivates our cascaded binary classification approach. Another challenge lies in the substantial overlap of MMSE scores between HC and Dementia groups, despite the expectation of clear separation. For instance, two Dementia subjects scored 29 out of 30, highlighting the subjectivity inherent in clinical assessments. Furthermore, MMSE scores are available for only 69 subjects, adding complexity to the regression task.

\vspace{-0.2cm}
\section{Methods}
\label{sec:Methods}
\vspace{-0.2cm}
This section outlines our methods for classification and regression tasks, covering data preprocessing, architectural details for both tasks (Figure~\ref{fig:Over workflow.}), and the ensemble strategies applied.
\vspace{-0.1cm}
\subsection{Preprocessing for classification and regression}
\vspace{-0.1cm}
% Preprocessing includes noise removal from the audio, application of ASR to the audio, and the incorporation of pause encoding in the transcripts.
\subsubsection{Denoising}
Through manual inspection of the dataset, we identified some audio recordings with loud background noise, beeping sounds at the start and end, and occasional interviewer interventions. To address the first two issues, we applied the MP-SENet model \cite{lu23e_interspeech}, which denoises both magnitude and phase spectra in parallel. Since interviewer interventions can negatively impact dementia detection performance \cite{liu2024leveraging}, we explored the use of an advanced speaker diarization model \cite{plaquet23_interspeech} to exclude the interviewer’s speech. However, initial experiments revealed frequent misclassifications, with substantial portions of the subject’s speech wrongly identified as the interviewer’s. Further comparisons between manual removal and retaining the interventions showed minimal impact on performance. Consequently, we opted to omit this step in subsequent study.
\vspace{-0.1cm}
\subsubsection{Fine-tuning ASR model}
\vspace{-0.1cm}
To enhance transcription accuracy for the test set, we fine-tune the advanced Automatic Speech Recognition (ASR) model, \textit{Whisper-large-v3}\footnote{https://huggingface.co/openai/whisper-large-v3}, using the challenge dataset. The 157 subjects are split into ASR training and validation sets in a 3:1 ratio, stratified by diagnostic labels, incorporating data from all three elicitation tasks for fine-tuning. Given the extended duration of the audio recordings, we first apply forced alignment \cite{yuan2008speaker} between each audio file and its corresponding manual transcript. Based on the alignment timestamps, we segment the audio and transcripts into 30-second audio-transcript pairs for fine-tuning.

Fine-tuning is conducted using the \textit{WhisperForConditionalGeneration} and \textit{WhisperProcessor} modules from the \textit{Transformers} Python library. The model checkpoint achieving the highest accuracy on the ASR validation set is selected for final transcription.
% This approach ensures precise synchronization between the audio and text.
\vspace{-0.1cm}
\subsubsection{Pause encoding}
\vspace{-0.1cm}
We encode pauses between words in input transcripts to better capture disfluency patterns. Specifically, transcripts are force-aligned with audio recordings using the same forced aligner \cite{yuan2008speaker}, where pauses are marked as ``SIL''. Pauses at the start and end of recordings are removed. Following \cite{yuan2020disfluencies}, pauses are categorized as short ($<$0.5s), medium (0.5–2s), or long ($>$2s), and encoded using “,”, “.”, and “...”, respectively. Since original punctuation is removed, these symbols exclusively represent pause information.
\vspace{-0.1cm}
\subsection{Classification task}
\vspace{-0.1cm}
\subsubsection{PLMs}
\vspace{-0.1cm}
Modern PLMs based on the Transformer architecture \cite{vaswani2017attention}, such as BERT\footnote{https://huggingface.co/google-bert/bert-large-uncased}, RoBERTa\footnote{https://huggingface.co/FacebookAI/roberta-large}, ERNIE\footnote{https://huggingface.co/nghuyong/ernie-2.0-large-en}, have proven capable of capturing a wide range of linguistic facts, including lexical knowledge, syntax, and semantics. We fine-tuned these three PLMs for dementia classification. Notably, we excluded speech-related features, as incorporating Pre-trained Speech Models (PSMs) such as Wav2Vec2 \cite{baevski2020wav2vec} or traditional handcrafted features did not yield noticeable performance gains.
\vspace{-0.1cm}
\subsubsection{Cascaded binary classification}
\vspace{-0.1cm}
Due to the significant data imbalance among diagnostic classes in the dataset and the inherent complexity of the three-class classification task, we propose a cascaded binary classification strategy to mitigate these challenges. Specifically, we replace the three-class classifier with two independently trained binary classifiers. As shown in Figure~\ref{fig:Over workflow.}, during the training phase, we first relabel all MCI and Dementia samples in the training set as Non-HC. We then train a binary classifier to differentiate HC from Non-HC using the entire training set (i.e., HC and Non-HC samples). This approach simplifies training, as the challenge dataset provides an approximate HC:MCI:Dementia ratio of 5:4:1, resulting in a nearly balanced HC:Non-HC distribution of approximately 1:1. Additionally, the distinction between Healthy Controls (HC) and patients (Non-HC) is likely more pronounced than the differences among the three original classes (HC, MCI, and Dementia).

For the second classifier, we train a binary model exclusively on MCI and Dementia samples. While this reduces the number of available training samples (excluding HC), the classification task is potentially less challenging than the original three-class classification, as it focuses solely on distinguishing between MCI and Dementia. 

During the testing phase, we first feed the preprocessed samples into an initial classifier to determine whether they belong to the HC or Non-HC. If a sample is classified as HC, it is directly assigned the HC label. Otherwise, it undergoes further processing by a second classifier to refine the classification into either MCI or Dementia. To construct these classifiers, we fine-tune PLMs. Specifically, for each preprocessed transcript, we input it into a PLM, such as BERT, and extract the 1024-dimensional embedding of the [CLS] token from the 24th layer. This embedding serves as input to a simple classification module comprising a linear layer with a Tanh activation function, a dropout layer, and a final linear layer with binary output.
% \vspace{-0.2cm}
\subsection{Regression task}
% Our preliminary experimental results indicate that when fine-tuning PLMs or PSMs for regression tasks—where a single linear output layer predicts the MMSE score—the predicted MMSE scores for individual samples tend to converge to the mean MMSE score of the training set, resulting in poor performance. Therefore, we adopt an alternative approach that leverages extracted multimodal features and employs machine learning regression models for training.
\subsubsection{Feature extraction}
We extract both acoustic and linguistic features from speech and text, respectively, to leverage complementary information across modalities. For acoustic features, we extract 1024-dimensional embeddings from the 24th transformer layer of Wav2Vec2, computed as the mean of all token embeddings. Additionally, we design a 10-dimensional silence feature set comprising the number of silence occurrences per second, the ratio of silence to speech duration, and statistical descriptors (maximum, minimum, mean, and standard deviation) of both silence and speech durations. Silence segments are identified using a Recurrent Neural Network (RNN)-based Voice Activity Detection (VAD) model from pyannote\footnote{https://github.com/pyannote/pyannote-audio}.

To capture low-level acoustic features, we employ the OpenSMILE toolkit\footnote{https://github.com/audeering/opensmile-python} to extract the eGeMAPS (88-dimensional) and ComParE (6373-dimensional) feature sets. For linguistic features, we obtain 1024-dimensional embeddings of the [CLS] token from the 24th transformer layer of RoBERTa. Consequently, we construct a total of five distinct feature sets.

\subsubsection{Regression models}
We use Support Vector Regression (SVR)\footnote{https://scikit-learn.org/stable/modules/generated/sklearn.svm.SVR.html}, XGBoost\footnote{https://github.com/dmlc/xgboost}, and CatBoost\footnote{https://github.com/catboost/catboost}, as existing research demonstrates their effectiveness in dementia detection \cite{mei2023ustc}.
\vspace{-0.2cm}
\subsection{Model ensemble}
\vspace{-0.2cm}
For both classification and regression tasks, we conduct experiments on each individual elicitation task. Each model undergoes 10 training runs with different random seeds to ensure robustness. In the classification task, for both Classifier 1 and Classifier 2, we employ majority voting across the output labels of three PLMs, three elicitation tasks, and 10 random seeds, resulting in a total of 3 × 3 × 10 = 90 models contributing to the final decision.

For the regression task, we first compute the average predictions across the 10 random seeds. Subsequently, we select models with a Root Mean Square Error (RMSE) below 3, drawn from three regression models applied to five feature sets across three elicitation tasks (3 × 5 × 3 = 45 models in total), and further refine the final MMSE score by averaging their outputs.

\section{Experiments}
\label{sec:Experiments}
\subsection{Experimental setup}
For the classification task, we partitioned the 157 subjects into training and validation sets in a 3:1 ratio based on diagnostic labels. To enhance robustness, we conducted four different data splits and evaluated performance using the Macro F1 score. Similarly, for the regression task, 69 subjects were divided into training and validation sets in a 3:1 ratio, maintaining the distribution of MMSE scores. We also performed four different splits and assessed model performance using RMSE.

The fine-tuning of PLMs was conducted over 20 epochs with a batch size of 8 and a learning rate of \(2 \times 10^{-5}\). We employed the AdamW optimizer and used cross-entropy loss for classification tasks. For regression models, optimal hyperparameters were determined via 4-fold cross-validation combined with a grid search strategy.

All experiments were conducted on NVIDIA RTX 4090 or A100 GPUs. Participants were allowed to submit up to three models for evaluation by the organizers.

\begin{table}[t!]
 \caption{Macro F1 score (\%) of the classification task.}
    \label{tab:Classification results}
    \centering
\resizebox{1.0\linewidth}{!}{
\begin{tabular}{cccc}
\toprule
\multirow{2}{*}{Different splits }           & \multicolumn{2}{c}{Validation set}      & \multirow{2}{*}{Test set}         \\
\cmidrule(lr){2-3} 

  & Cascaded binary   & Direct three-class  &  \\
\midrule
Split 1            & 59.9              & 55.3                &    $-$           \\
Split 2            & 57.9              & 57.3                &    $-$     \\
Split 3            & 63.6 \makebox[-1.5pt][l]{\checkmark} & 59.9 & 45.1           \\         
Split 4            & 61.8              & 61.1                &    $-$             \\
% \midrule
\multicolumn{3}{c}{Majority voting (4 splits)\makebox[-1.5pt][l]{\checkmark}}           & 57.8            \\
% \midrule
\multicolumn{3}{c}{Majority voting (Retraining(4 splits))\makebox[-1.5pt][l]{\checkmark}}  &\textbf{58.6}          \\
\midrule
\midrule
\multicolumn{3}{c}{Baseline}                                 & 55.0            \\
\bottomrule
\end{tabular}}
\vspace{-0.6cm}
\end{table}
% \vspace{-0.2cm}
\subsection{Results and analyses}
\subsubsection{ASR model performance}
After fine-tuning, the Word Error Rate (WER) of Whisper on the ASR validation set decreased from the original 29.7\% to 23.2\%, demonstrating the effectiveness of the method.
\vspace{-0.1cm}
\subsubsection{Classification task}
\vspace{-0.1cm}
For the classification task, we conducted experiments by fine-tuning PLMs for direct three-class classification, serving as a baseline for comparison with the proposed cascaded binary classification approach. The three submitted models were selected based on the following criteria: (1) the model that achieved the highest validation performance among the four data splits; (2) an ensemble model leveraging majority voting across all models from the four splits; and (3) a model obtained by retraining all models using the entire dataset, followed by majority voting to aggregate their predictions.

Table~\ref{tab:Classification results} presents the performance comparison between the proposed cascaded binary classification approach and the direct three-class classification method, both employing model ensemble, on the validation set of each split. Additionally, it reports the performance of the selected models on the test set based on the predefined criteria. From the results, we observe the following key findings:

\textit{\textbf{1)}} The proposed cascaded binary classification approach consistently outperforms direct three-class classification on the validation set, effectively simplifying multiclass classification and mitigating data imbalance across diagnostic classes. \textit{\textbf{2)}} On the test set, our method achieved an F1 score of 58.6\%, exceeding the baseline (55\%) by 3.6\%, further confirming its effectiveness. \textit{\textbf{3)}} Although Split 3 showed strong validation performance, it suffered a significant drop on the test set, indicating potential overfitting. However, a multi-split ensemble strategy improved the test F1 score, enhancing model robustness. \textit{\textbf{4)}} Finally, retraining on the entire dataset yielded further performance gains, underscoring the advantages of full data utilization for model optimization.

% (a) The proposed cascaded binary classification method consistently outperformed the direct three-class classification approach on the validation set. This suggests that our method effectively reduces the complexity of multiclass classification while mitigating the substantial data imbalance across diagnostic classes.
% (b) On the test set, our approach achieved an F1 score of 58.6\%, surpassing the baseline (55\%) by 3.6\%, further validating its effectiveness.
% (c) While Split 3 exhibited strong performance on the validation set, it experienced a notable performance drop on the test set, indicating potential overfitting. However, employing a multi-split ensemble strategy improved the F1 score on the test set, highlighting its role in enhancing model robustness.
% (d) Retraining the model on the entire dataset led to additional performance gains, demonstrating the benefits of leveraging the full dataset for model optimization.
\vspace{-0.1cm}
\subsubsection{Regression task}
\vspace{-0.1cm}
For the regression task, we explored fine-tuning RoBERTa and Wav2Vec2 by appending a single-output linear layer to each model and averaging their outputs to predict the MMSE score. This approach was compared against our proposed method, which extracts multimodal features and integrates them with machine learning regression models. The selection criteria for the three submitted models were aligned with those used in the classification task: (1) selecting the model from the split that achieved the highest validation performance among the four splits; (2) averaging the scores from all models across the four splits; and (3) retraining all models using the entire dataset and subsequently averaging their predicted scores.

Table 2 presents the performance of the proposed multimodal feature fusion method alongside fine-tuned RoBERTa and Wav2Vec2 models with ensemble learning, evaluated on the validation set of each split and the test set for selected models. Key observations include:

\textit{\textbf{1)}} The proposed multimodal feature fusion method consistently outperforms fine-tuned pre-trained models in terms of RMSE on the validation set. Fine-tuned models often predict MMSE scores near the mean of the training data, likely due to the dataset's limited size (69 subjects). The complexity and larger parameter size of pre-trained models make it challenging to capture dementia-specific features from such a small dataset. This highlights the advantage of multimodal feature extraction combined with simpler machine learning models for regression tasks.
\textit{\textbf{2)}} All three submitted models surpassed the baseline (RMSE = 2.98), with the best achieving an RMSE of 2.87, an 0.11 improvement, demonstrating the effectiveness of our proposed approach for MMSE score prediction.
\textit{\textbf{3)}} Model ensembles integrating diverse acoustic and linguistic features consistently achieved superior performance across all splits, underscoring the complementary nature of these modalities and validating the multimodal fusion framework.
\textit{\textbf{4)}} Despite these gains, the model exhibited mild overfitting on the validation set, indicating the need for stronger generalization strategies. Additionally, dataset-specific challenges, such as overlapping MMSE scores across diagnostic categories and class imbalance (Section~\ref{sec:Dataset}), further complicate model optimization.

\begin{table}[t!]
 \caption{RMSE of the regression task.}
    \label{tab:Regression results}
    \centering
\resizebox{1.0\linewidth}{!}{
\begin{tabular}{cccc}
\toprule
\multirow{2}{*}{Different splits }           & \multicolumn{2}{c}{Validation set}      & \multirow{2}{*}{Test set}         \\
\cmidrule(lr){2-3} 

  &        Multimodal   & Fine-tuning RoB+W2V &  \\
\midrule
Split 1            & 2.88              & 2.95                &    $-$         \\
Split 2            & 2.56 \makebox[-1.5pt][l]{\checkmark}  & 2.83   &    2.88          \\
Split 3            & 2.66  & 2.98 & $-$          \\         
Split 4            & 2.92              & 2.97                &    $-$             \\
% \midrule
\multicolumn{3}{c}{Score averaging  (4 splits)\makebox[-1.5pt][l]{\checkmark}}           & \textbf{2.87}          \\
% \midrule
\multicolumn{3}{c}{Score averaging  (Retraining(4 splits))\makebox[-1.5pt][l]{\checkmark}}  &2.89          \\
\midrule
\midrule
\multicolumn{3}{c}{Baseline}                                 & 2.98            \\
\bottomrule
\end{tabular}}
\vspace{-0.6cm}
\end{table}

\vspace{-0.2cm}
\section{Conclusions}
\vspace{-0.1cm}
\label{sec:Conclusions}
In this paper, we present the results and analyses of our submission to this year’s PROCESS challenge. For the classification task, we introduce a cascaded binary classification strategy to tackle the underexplored three-class problem, with particular emphasis on the early diagnosis of MCI. This approach simplifies multi-class classification and mitigates data imbalance across diagnostic classes. For the regression task, we propose an enhanced multimodal framework that integrates diverse acoustic and linguistic features, leveraging their complementary strengths for improved performance. Experimental results on the test set demonstrate that our methods outperform the provided baseline in both tasks, underscoring the potential of our approach for advancing early dementia detection. Future work will focus on developing more robust and generalizable automated dementia detection models for real-world applications.
\section{Acknowledgements}
This work was partially supported by the National Social Science Foundation of China (Grant No. 23AYY012), and by the Supercomputing Center of the University of Science and Technology of China.
\bibliographystyle{IEEEtran}
\bibliography{mybib}

\end{document}